\documentclass[11pt]{article}
\usepackage{graphicx}
\usepackage{placeins}
\usepackage{float}
\usepackage{subfigure}
\usepackage[toc,page]{appendix}
\usepackage{amssymb}
\usepackage{epsfig}
\begin{document}
\title{{\bf{\Large Free-fall energy density and flux in the Schwarzschild black hole }}}
\author{ 
{\bf {\normalsize Wontae Kim $^{1}$}$
$\thanks{E-mail: wtkim@sogang.ac.kr}} and
{\bf {\normalsize Bibhas Ranjan Majhi$^{2}$}$
$\thanks{E-mail: bibhas.majhi@mail.huji.ac.il}
}$~$
\\\\$~$
{\normalsize $^1$ Department of Physics, Sogang University, Seoul 121-742, Republic of Korea}
\\$~$
{\normalsize $^2$ Racah Institute of Physics, Hebrew University of Jerusalem,}
\\{\normalsize Givat Ram, Jerusalem 91904, Israel}
\\[0.3cm]
}
\maketitle

\begin{abstract}
In the four-dimensional background of Schwarzschild black hole, 
we investigate the energy densities and fluxes in the freely falling frames
for the Boulware, Unruh, and Israel-Hartle-Hawking states.
In particular, we study their behaviors near the horizon
and asymptotic spatial infinity by using the 
trace anomaly of a conformally invariant scalar field.
In the Boulware state, both the energy density and flux are negative divergent 
when the observer is dropped at the horizon, and asymptotically vanish. In the Unruh state, 
the energy density is also negative divergent at the horizon while it is positive finite
asymptotically. The flux in the Unruh state is always positive
and divergent at the horizon. 
In the Israel-Hartle-Hawking state, the energy density depends on
the angular motion of free fall,
and fluxes vanish at the horizon and the spatial infinity. 
Finally, we discuss the role of the negative energy density  
near the horizon in the evaporating black hole. 
\end{abstract}

\newpage
\section{Introduction}
Black holes are still mysterious objects in that they can be excited thermally like a black body; however, they can evaporate eventually through Hawking radiation \cite{Hawking:1974sw,Hawking:1976ra}, 
which yields information loss problem and related 
intriguing issues \cite{Susskind:1993if,Stephens:1993an,Susskind:1993mu}. The original particle explanation of Hawking radiation is based on quantum-mechanical pair creations consisting of the positive and negative energy states
with appropriate boundary conditions. 
If Hawking radiation plays a role of  information carrier, then the full information can be
recovered by accumulating the whole radiation. Thus it is expected that the quantum-mechanical unitary evolution of black hole
will be accommodated. Of course, there is another way 
to look into some aspects of Hawking radiation, which can be 
realized near the horizon along the line of Unruh \cite{Unruh:1976db}. 
At the horizon, it has been believed that the Kruskal coordinates play a role
of free-fall coordinate and the space-time is locally flat. In that region, there 
are no out-going modes and
the space-time is regarded as a free-fall vacuum where there is no radiation,
whereas the fixed observer in the Schwarzschild coordinates 
which is equivalent to the accelerated observer in the Rindler coordinates
can perceive radiation following the Unruh effect \cite{Unruh:1976db}.

On the other hand, 
there is an interesting observation which is of relevance to  
a constant negative energy density that was measured by an observer orbiting close to a black hole
 \cite{Ford:1993bw}. Now, one can extend 
 this to a time-like radial geodesic curve as a trajectory of freely falling
observer and study the free-fall energy density around a black hole.
In the collapsing two-dimension dilatonic black hole \cite{Callan:1992rs}, 
the negative energy density was found
near the horizon which is divergent at the horizon when the free fall 
begins just at the horizon \cite{Kim:2013caa}. It has also 
been shown that the positive and negative energy regions are 
separated by a time like curve and the negative energy region 
is reduced as time goes on.  
This calculation can also be done in the two-dimensional soluble Schwartzschild black hole
for three black hole states of the Boulware \cite{Boulware:1404}, Unruh \cite{Unruh:1976db}, and Israel-Hartle-Hawking
states \cite{Israel:1976ur,Hartle:1976tp} by employing
the trace anomaly of scalar fields \cite{Christensen:1977jc}.
It turns out that the behaviors of the energy-density rely on
the initial free-fall status \cite{Eune:2014eka}
in the sense that  
the negative energy density is dominant near the horizon
and is divergent just at the horizon but it can be surprisingly positive finite
even at the horizon in the Unruh state depending on the starting position of the free fall. 
The stretched horizon observed by the freely falling observer was also discussed in
connection with the negative energy density related to instability of black hole
\cite{Park:2014mba}. And, the detector response has been studied for geodesic trajectories 
in the context of Hawking phenomenon \cite{Suprit} which predicts higher temperature 
near the horizon. This implies that the energy flux near the horizon measured by these 
observers must diverge.  Recently, the negative energy region was defined as
``the zone''  \cite{Freivogel:2014dca} and its role was 
intensively discussed in connection with
the firewall argument \cite{Almheiri:2012rt} (for a similar prediction from different assumptions, see~\cite{Braunstein:2013bra}).
On the other hand, it has also been suggested that 
there is no apparent need for firewalls since the unitary evolution of black hole 
entangles a late mode located outside the event horizon with a combination 
of early radiation and black hole states \cite{Hutchinson:2013kka},
and claimed that the remaining set of nonsingular realistic states do not have 
firewalls but yet preserve information \cite{Page:2013mqa}.

In this work, we would like to study the behaviors of the energy density and flux measured
by the freely falling observer on the Schwarzschild black hole background. 
The advantage of lower dimensional models 
was to be able to determine the energy-momentum tensors exactly by solving 
trace anomaly equation and covariant conservation relation. However, 
in the four dimensional black hole it is non-trivial task to obtain
 the exact form of the energy-momentum
tensors from the conformal anomaly  analytically \cite{Christensen:1977jc,Candelas}.
Fortunately, the essential ingredient in black holes mostly
comes from the near horizon and the asymptotic region where the complicated
equations are relatively simplified \cite{Balbinot:1999vg}. 
So we are going to calculate the energy density and the flux for three
different black hole states in the free-fall frame, and study 
what state is relevant to the angular free-fall motion at the event
horizon and examine whether the similar divergent behavior of energy density
at the horizon appears or not in the four-dimensional case. This is very important
 because this negative divergent behavior of the energy density 
 is not the two-dimensional characteristics but more or less generic feature.

The organization of the paper is as follows.
In section 2, the  
proper velocity from the geodesic equation is obtained 
and the radial velocity is chosen as zero for convenience 
when the free-fall begins toward the black hole. 
In the next section, the approximate energy-momentum tensors near 
the horizon and the spatial infinity are introduced for three black hole states in the self-contained
manner \cite{Balbinot:1999vg}.
In section 4, it will be shown that in the Boulware state the energy density is negative divergent 
at the horizon, and asymptotically vanishes. In the Unruh state, 
it is negative at the horizon whereas it is positive finite
asymptotically. For the Boulware and the Unruh states, the angular motion 
of free fall is not significant at the horizon.
However, in the Israel-Hartle-Hawking state, 
for $2M > L$, the energy density is negative at the horizon
whereas for $ 2M < L $ it can be positive.
The behaviour of the flux, near the horizon and at the spatial infinity,
 will be discussed in section $5$. 
The final section is devoted to conclusion and discussion.

\section{\label{sec2}The setup: Freely falling observer in black hole spacetime}
In this section, the components of the velocity vector for a freely falling
       observer in the Schwarzschild spacetime will be given. These are found out by solving geodesic equations.  
The four dimensional Schwarzschild black hole space time 
is given by the following metric,
\begin{equation}
ds^2 = -f(r)dt^2+\frac{dr^2}{f(r)}+r^2\Big(d\theta^2 + \sin^2\theta d\phi^2\Big)~,
\label{metric} 
\end{equation}
where the metric function is $f(r) = 1-2M/r$ and the event horizon is located at $r_H = 2M$. The metric coefficients are independent of time $t$ and the angular
$\phi$ coordinates. They correspond to two conserved quantities, namely 
the energy and angular momentum. Moreover, since the metric (\ref{metric}) is spherically symmetric, the motion of the observer is confined on the plane. Therefore, there is no motion along the normal to the plane and hence $\theta=\pi/2$ is
chosen.  
Using this fact, it is easy to solve geodesic equations. Remember that in the subsequent analysis, without any loss of generality, we will consider $\theta=\pi/2$ (For more details, see section $7.4$ of \cite{Padmanabhan:2010zzb}).
 Solution of geodesic equation leads to the following form of the proper velocity of the freely falling observer,   
\begin{equation}
u^a = \Big(\frac{dt}{d\tau},\frac{dr}{d\tau},\frac{d\theta}{d\tau},\frac{d\phi}{d\tau}\Big) = \Big(\frac{E}
{f(r)},\pm[E^2 - V_{eff}^2(r)]^{1/2},0,\frac{L}{r^2}\Big) 
\label{velocity}
\end{equation}
where 
\begin{equation}
V_{eff}^2(r) = f(r)\Big(1+\frac{L^2}{r^2}\Big)~.
\label{V} 
\end{equation}
Here, $E$ and $L$ are two constants which are conserved quantities corresponding to time translational and rotational symmetry of the spacetime metric (\ref{metric}), respectively. They are identified with the energy and angular momentum per unit mass. 
The positive sign refers to the outgoing observer away from the horizon 
and the negative sign corresponds to the ingoing observer toward the horizon. 
In our calculations, we shall consider only the ingoing observer. 

   Now, it can be shown that the energy and angular momentum are related by the initial position of the observer, 
   provided the initial velocity is zero. It will be done in the following way. The radial velocity of the observer with respect to the Schwarzschild time is 
\begin{equation}
\Big(\frac{dr}{dt}\Big)^2 = \Big(\frac{f(r)}{E}\Big)^2 \Big[E^2 - V_{eff}^2(r)\Big]~.
\label{Schvelo} 
\end{equation}
Note that the radial velocity always vanishes at $r=2M$, which is independent of
the energy. 
If the initial velocity of the frame is zero at $r_s$, 
then we obtain $E^2 -V_{eff}^2(r_s) = 0$, 
where $r_s$ is the initial position of the observer.
This helps us to express 
the conserved quantity $E$ as
\begin{equation}
E = V_{eff}(r_s) = \sqrt{f(r_s)\Big(1+\frac{L^2}{r^2_s}\Big)},
\label{E} 
\end{equation}
which can be reduced to the two-dimensional case for
$L=0$ \cite{Eune:2014eka}. 
This will be used later to express the energy density and flux 
in terms of the initial position of the observer.

\section{Components of renormalized energy-momentum tensor}
  In four dimensions, breaking of 
  the conformal symmetry leads to the trace anomaly, and 
  the form of the anomalous energy-momentum tensor can be obtained by solving two equations: conservation of the stress-tensor and trace anomaly expression
  \cite{Christensen:1977jc}.
   Unfortunately, the exact form of the components can not be evaluated. 
   But still, it is possible to evaluate only the asymptotic forms near the horizon and at infinity which are necessary for our analysis. The value of the integration constants appearing in the solutions, depends on the nature of the vacuum. In the below, we give the components in ($t,r,\theta,\phi$) coordinates for three vacua. Near the horizon ($r\rightarrow 2M$) these are given by \cite{Christensen:1977jc, Candelas,Balbinot:1999vg}
\begin{equation}
\begin{array}{lc}
  \Big<T^a_b\Big>_B \rightarrow C^{(0)}_B& \left(\begin{array}{@{}cccc@{}}
                    -1 & 0 & 0 & 0 \\
                     0 & 1/3 & 0 & 0 \\
                    0 & 0 & 1/3 & 0 \\
                    0 & 0 & 0 & 1/3
                  \end{array}\right) \\[15pt]
\end{array},
\end{equation}
\begin{equation}
\begin{array}{lc}
  \Big<T^a_b\Big>_U \rightarrow C^{(0)}_U& \left(\begin{array}{@{}cccc@{}}
                    1/f & -1 & 0 & 0 \\
                     1/f^2 & -1/f & 0 & 0 \\
                    0 & 0 & 0 & 0 \\
                    0 & 0 & 0 & 0
                  \end{array}\right) \\[15pt]
\end{array},
\end{equation}
\begin{equation}
\begin{array}{lc} \label{Thh}
  \Big<T^a_b\Big>_H \rightarrow C^{(0)}_H& \left(\begin{array}{@{}cccc@{}}
                    1 & 0 & 0 & 0 \\
                     0 & 1 & 0 & 0 \\
                    0 & 0 & 2 & 0 \\
                    0 & 0 & 0 & 2
                  \end{array}\right) \\[15pt]
\end{array},
\end{equation}
where in the above $B, U$ and $H$ stand for the Boulware state \cite{Boulware:1404}, 
Unruh state \cite{Unruh:1976db}, and Israel-Hartle-Hawking state  \cite{Israel:1976ur,Hartle:1976tp}, respectively. 
The expressions for $C^{(0)}$'s are
\begin{equation}
C^{(0)}_B = -\frac{1}{30~ 2^{12}\pi^2 M^4 f^2}, \,\,\,\ C^{(0)}_U = \frac{\mathcal{L}}{4\pi(2M)^2}, \,\,\,\ 
C^{(0)}_H = \frac{1}{96 M^4},
\label{CH} 
\end{equation}
where $\mathcal{L}$ is called the Luminosity which is always positive and finite quantity \cite{Balbinot:1999vg}. 
On the other hand, the energy-momentum tensor component 
in $r\rightarrow \infty$ region are
\begin{equation}
\begin{array}{lc}
  \Big<T^a_b\Big>_B \rightarrow & \begin{array}{@{}c@{}}
                    \mathcal{O}(r^{-6}) 
                  \end{array} \\[15pt]
\end{array},
\end{equation}
\begin{equation}
\begin{array}{lc}
  \Big<T^a_b\Big>_U \rightarrow C^{(\infty)}_U& \left(\begin{array}{@{}cccc@{}}
                    -1 & -1 & 0 & 0 \\
                     1 & 1 & 0 & 0 \\
                    0 & 0 & 0 & 0 \\
                    0 & 0 & 0 & 0
                  \end{array}\right) \\[15pt]
\end{array},
\end{equation}
\begin{equation}
\begin{array}{lc}
  \Big<T^a_b\Big>_H \rightarrow C^{(\infty)}_H& \left(\begin{array}{@{}cccc@{}}
                    -1 & 0 & 0 & 0 \\
                     0 & 1/3 & 0 & 0 \\
                    0 & 0 & 1/3 & 0 \\
                    0 & 0 & 0 & 1/3
                  \end{array}\right) \\[15pt]
\end{array},
\end{equation}
where
\begin{equation}
C^{(\infty)}_U = \frac{\mathcal{L}}{4\pi r^2}, \,\,\,\ 
C^{(\infty)}_H = \frac{1}{30~ 2^{12}\pi^2 M^4}~.
\label{Cinf} 
\end{equation} 
It must be noted that the above expressions are valid only in the two regions 
and reflects the asymptotic nature of them, not the exact form. 
Hence any physical quantity obtained from these components is also approximate. But interesting thing is that we can still say about the nature of the quantities which are our interest of study. In the below, we shall discuss the energy density and flux measured by a freely falling observer. 

\section{Energy densities in three different states}
The energy density for a freely falling observer in different states can be calculated using the following expression 
\begin{equation}
\epsilon = \Big<T^a_b\Big> u_au^b~.
\label{energy}
\end{equation}
In this calculation, the results are valid up to the leading order in $1/f(r)$ 
for the case of near horizon limit or the inverse of radial distance $r$ for
the case of the 
asymptotic infinite limit. The other terms are irrelevant 
because we are only interested in the asymptotic behavior of the
energy density. 

Let us first start with the Boulware state.
In the region of $r\rightarrow 2M$ along with $\theta = \pi/2$, it can be given as
\begin{eqnarray}
\epsilon_{B}^{(0)}(r|r_s) &=& C^{(0)}_B \Big[\Big<T^t_t\Big>u_tu^t + \Big<T^r_r\Big>u_ru^r + \Big<T^\phi_\phi\Big>u_\phi 
u^\phi\Big]
\nonumber
\\
&=& C^{(0)}_B \Big[\frac{E^2}{f(r)} + \frac{1}{3f(r)}(E^2 - V_{eff}^2(r))+\frac{L^2}{3r^2}\Big].
\label{eHB}
\end{eqnarray}
Using Eqs. (\ref{V}) and (\ref{E}), the expression (\ref{eHB}) 
can be simplified as
\begin{equation}
\epsilon^{(0)}_B(r|r_s) = -\frac{1}{30 ~2^{12}\pi^2M^4f^2(r)}\Big[\frac{4f(r_s)}{3f(r)}\Big(1+\frac{L^2}{r^2_s}\Big)
-\frac{1}{3}\Big]~, 
\end{equation}
which is always negative divergent at the horizon. 
Now if the measurement is done at the starting point of free fall, 
we can obtain
\begin{equation}
\epsilon^{(0)}_B(r_s|r_s) = -\frac{1}{30 ~2^{12}\pi^2M^4f^2(r_s)}\Big(1+\frac{4L^2}{3r^2_s}\Big)~. 
\end{equation}
This tells that the freely falling observer, which begins its journey very close to the horizon, sees very large negative energy density. For other limit $r\rightarrow\infty$, the energy density falls as
$\epsilon_{B}^{(\infty)} \rightarrow \mathcal{O}(1/r^6)$.

For the case of the Unruh state, the energy density near the horizon is given by
\begin{eqnarray}
\epsilon^{(0)}_U(r|r_s) &=& C^{(0)}_U \Big[\Big<T^t_t\Big>u_tu^t + \Big<T^t_r\Big>u_tu^r + \Big<T^r_t\Big>u_ru^t + 
\Big<T^r_r\Big>u_ru^r\Big] 
\nonumber
\\
&=& C^{(0)}_U \Big[-\frac{E^2}{f^2(r)}+ \frac{2E}{f^2(r)} 
\Big(E^2 - V_{eff}^2(r)\Big)^{1/2} 
- \frac{1}{f^2(r)}\Big(E^2 - V_{eff}^2(r)\Big)\Big]~.
\label{eHU} 
\end{eqnarray}
Using Eqs. (\ref{V}) and (\ref{E}), we can obtain
\begin{eqnarray}
\epsilon^{(0)}_U(r|r_s) &=&  C^{(0)}_U\Big[-\frac{2f(r_s)}{f^2(r)}\Big(1+\frac{L^2}{r_s^2}\Big) + \frac{2\sqrt{f(r_s)\Big(1+\frac{L^2}{r_s^2}\Big)}}{f^2(r)}\Big\{f(r_s)\Big(1+\frac{L^2}{r_s^2}\Big)
\nonumber
\\
&-& f(r)\Big(1+\frac{L^2}{r^2}\Big)\Big\}^{1/2}+\frac{1+\frac{L^2}{r^2}}{f(r)}\Big]~.
\end{eqnarray}
The energy density measured at the moment when the free fall begins 
is reduced to 
\begin{equation}
\epsilon^{(0)}_U(r_s|r_s) =  -\frac{\mathcal{L}}{4\pi(2M)^2} \frac{\Big(1+\frac{L^2}{r_s^2}\Big)}{f(r_s)}~.
\end{equation}
Since luminosity is positive where its numerical value is $\mathcal{L} = 
2.337\times 10^{-4}/(\pi M^2)$ \cite{Elster}, 
the energy density is always negative divergent for $r_s\rightarrow 2M$.
In the asymptotically infinity of $r\rightarrow\infty$, it turns out to be
\begin{eqnarray}
\epsilon^{(\infty)}_U(r|r_s) &=& C^{(\infty)}_U \Big[\Big<T^t_t\Big>u_tu^t + \Big<T^t_r\Big>u_tu^r + \Big<T^r_t\Big>u_ru^t + 
\Big<T^r_r\Big>u_ru^r\Big] 
\nonumber
\\
&=& C^{(\infty)}_U \Big[\frac{E^2}{f(r)} + E\Big(E^2 - V_{eff}^2(r)\Big)^{1/2} + \frac{E}{f^2(r)} 
\Big(E^2 - V_{eff}^2(r)\Big)^{1/2} 
\nonumber
\\
&+& \frac{1}{f(r)}\Big(E^2 - V_{eff}^2(r)\Big)\Big]~.
\label{einfU} 
\end{eqnarray}
Now, if we impose the condition that 
the observer's initial starting point and measurement point are the same,
then it leads to
\begin{equation}
\epsilon^{(\infty)}_U(r_s|r_s) =   C^{(\infty)}_U \Big(1+\frac{L^2}{r_s^2}\Big)~.
\end{equation}
Note that it is positive and vanishes in the limit $r_s\rightarrow\infty$. 
But remember that this is the energy density. To obtain the energy,
 we need to integrate Eq. (\ref{einfU}) over the three space volume and then take the limit $r_s\rightarrow\infty$. 
This will lead to a finite quantity which is proportional to squire of the Hawking temperature $i.e.$, $T_H^2 = (1/8\pi M)^2$, since luminosity is proportional to 
the inverse squire of the mass $M$ of the black hole. In the two dimensional case, the energy density at infinity is itself finite \cite{Eune:2014eka}. This is because there is no angular part contribution in the stress-tensor, whereas in the four-dimensional
case the angular part provides the $1/r^2$ dependence in $C_U^{(\infty)}$, as seen 
from Eq. (\ref{Cinf}), which spoils the finite nature in the infinity limit. 
In other words, in the two-dimensional case
 there is no angular integration to obtain the energy, which means 
 the energy density itself must have
  the finiteness. So there 
is no discrepancy between them. 

Finally,
 we discuss the nature of the energy density in the Israel-Hartle-Hawking state.
For $r\rightarrow 2M$, it is given as
\begin{equation}
\epsilon^{(0)}_H(r|r_s) = C^{(0)}_H \Big[-\frac{V_{eff}^2(r)}{f(r)} + \frac{2L^2}{r^2}\Big]~
\label{eHH} 
\end{equation}
and using Eqs. (\ref{V}) and (\ref{E}) in Eq. (\ref{eHH}) leads to
\begin{equation}
\epsilon^{(0)}_{H}(r|r_s) = -\frac{1}{96M^4} \Big(1-\frac{L^2}{r^2}\Big)~. 
\label{1.01}
\end{equation}
Note that Eq. (\ref{1.01}) is insensitive to the initial position of the observer
in the near horizon limit so that
the initial free-fall position in this limit does not alter the essential behavior of the energy
density. This is also compatible with the
two-dimensional calculations in the near horizon limit of
$L=0$ where it is always negative \cite{Eune:2014eka}.  
The crucial difference from the two-dimensional case 
comes from the fact that for a given non-vanishing $L$
 there exists additional positive contribution to the
  energy density.
For $2M > L$, the energy density (\ref{1.01})
can be still negative even at the horizon
whereas for $ 2M < L $ it can be positive at the horizon.
For the latter case, there is also 
a critical free-fall position  of $r_c=L > r_H$ where
the energy density vanishes. It means that the 
highly circular free-fall motion near the horizon gives rise to  
transition from the negative energy density to the positive energy density
during free fall.
On the other hand, for $r\rightarrow\infty$, in the similar manner 
we can obtain
\begin{equation}
\epsilon^{(\infty)}_H (r|r_s) = C^{(\infty)}_H \Big[\frac{E^2}{f(r)} + \frac{1}{3f(r)}\Big(E^2 - V_{eff}^2(r)\Big)+
\frac{L^2}{3r^2}\Big]~,
\label{einfH} 
\end{equation}
where $C^{(\infty)}_H$ is finite so that 
the energy density fall as $\mathcal{O}(1/r^2)$. For $r=r_s$ the above 
equation is simply reduced to
\begin{equation}
\epsilon^{(\infty)}_H(r_s|r_s) = C^{(\infty)}_H \Big[1+\frac{4L^2}{3r_s^2}\Big]~, 
\end{equation}
which is definitely positive.
\section{Fluxes in different states}
In this section, we shall study the asymptotic behavior of the flux 
for a freely falling observer in three states. 
The flux for a freely falling observer is defined as \cite{Ford:1993bw}
\begin{equation}
F = -\Big<T^a_b\Big>u_an^b
\label{flux} 
\end{equation}
where $n^a$ is a spacelike unit normal satisfying $g_{ab}n^an^b=+1$ 
and $g_{ab}n^au^b=0$.
In order to determine $n^a$, let us suppose that
 the normal vector has the form of $n^a=(A,B,C,D)$ where the unknown parameters
 have to be fixed by using two conditions 
$n_an^a=+1$ and $n_au^a=0$. Since the unknown parameters are four while the number of equations relating them are two, 
we impose the condition $C=0=D$ from the beginning for convenience to eliminate 
the redundant degrees of freedom. Then, these two conditions yield
\begin{eqnarray}
&&f(r)A^2 - \frac{B^2}{f(r)}=1~,
\\
&&AE +\frac{B}{f(r)}(E^2 - V_{eff}^2(r))^{1/2}=0~.
\end{eqnarray}
Solving the above two equations, we obtain
\begin{equation}
A= \mp\frac{\Big(E^2-V_{eff}^2(r)\Big)^{1/2}}{V_{eff}(r)\sqrt{f(r)}}, \,\,\,\ B=\pm\frac{E\sqrt{f(r)}}{V_{eff}(r)}~. 
\end{equation}
As we mentioned earlier, the negative sign in Eq. (\ref{velocity}) has been chosen.
Now considering the positive sign of $B$, we find the components of the unit normal in the case of the present metric  
(\ref{metric}) as
\begin{equation}
n^a = \Big(-\frac{\Big(E^2-V_{eff}^2(r)\Big)^{1/2}}{V_{eff}(r)\sqrt{f(r)}},\frac{E\sqrt{f(r)}}{V_{eff}(r)},0,0\Big)~. 
\end{equation}
The expressions for flux measured by a freely falling observer 
will be given for three 
different states for the regions $r\rightarrow 2M$ and $r\rightarrow \infty$ 
in what follows.

For the Boulware state, 
the value of the flux for the limit $r\rightarrow 2M$ turns out to be
\begin{eqnarray}
F^{(0)}_B (r|r_s) &=& -\frac{1}{30~2^{12}\pi^2M^4}\frac{4E\Big(E^2-V_{eff}^2(r)\Big)^{1/2}}{3V_{eff}(r)f^{5/2}(r)} 
\nonumber
\\
&=&  -\frac{1}{30~2^{12}\pi^2M^4}\frac{4f(r_s)\Big(1+\frac{L^2}{r_s^2}\Big)\Big[1-\frac{f(r)\Big(1+\frac{L^2}{r^2}\Big)}{f(r_s)\Big(1+\frac{L^2}{r_s^2}\Big)}\Big]^{1/2}}{3f^{3}(r)\sqrt{\Big(1+\frac{L^2}{r^2}\Big)}} 
\end{eqnarray}
which diverges near the horizon. Moreover, the above vanishes for $r=r_s$.  On the other hand, this falls as $F^{(\infty)}_B (r|r_s)\rightarrow\mathcal{O}(r^{-6})$ for
the limit of $r\rightarrow\infty$.

 In the case of the Unruh state, 
 the flux for the near horizon limit is given by
\begin{eqnarray}
F^{(0)}_U(r|r_s) &=& -\frac{\mathcal{L}}{4\pi(2M)^2}\Big[\frac{2E\Big(E^2-V_{eff}^2(r)\Big)^{1/2}}{f^{3/2}(r)V_{eff}(r)}-\frac{E^2}{f^{3/2}(r)V_{eff}(r)}-\frac{E^2-V_{eff}^2(r)}{f^{3/2}(r)V_{eff}(r)}\Big]
\nonumber   
\\
&=& \frac{\mathcal{L}}{4\pi(2M)^2}\Big[-\frac{2f(r_s)\Big(1+\frac{L^2}{r_s^2}\Big)\Big(1-\frac{f(r)\Big(1+\frac{L^2}{r^2}\Big)}{f(r_s)\Big(1+\frac{L^2}{r_s^2}\Big)}\Big)^{1/2}}{f^{2}(r)\Big(1+\frac{L^2}{r^2}\Big)^{1/2}} + f(r_s)\frac{\Big(1+\frac{L^2}{r_s^2}\Big)}{f^2(r)\Big(1+\frac{L^2}{r^2}\Big)^{1/2}}
\nonumber
\\
&& + \frac{f(r_s)\Big(1+\frac{L^2}{r_s^2}\Big)\Big(1-\frac{f(r)\Big(1+\frac{L^2}{r^2}\Big)}{f(r_s)\Big(1+\frac{L^2}{r_s^2}\Big)}\Big)}{f^{2}(r)\Big(1+\frac{L^2}{r^2}\Big)^{1/2}}\Big]
\end{eqnarray}
and in particular at $r=r_s$ it can be simplified as 
\begin{equation}
F^{(0)}_U(r_s|r_s) =  \frac{\mathcal{L}}{4\pi(2M)^2} \frac{1}{f(r_s)}\Big(1+\frac{L^2}{r_s^2}\Big)^{1/2}~.
\end{equation}
Since luminosity is positive finite quantity, it is always positive and finite except at $r_s=2M$ where it diverges.
In the asymptotic infinity region, the flux for the Unruh vacuum is calculated as
\begin{eqnarray}
F^{(\infty)}_U(r|r_s) &=& \frac{\mathcal{L}}{4\pi r^2}\Big[\frac{2E\Big(E^2-V_{eff}^2(r)\Big)^{1/2}}{\sqrt{f(r)}V_{eff}(r)}+\frac{E^2\sqrt{f(r)}}{V_{eff}(r)}+ \frac{E^2-V_{eff}^2(r)}{f^{3/2}(r)V_{eff}(r)}\Big]
\nonumber
\\
&=& \frac{\mathcal{L}}{4\pi r^2}\Big[\frac{2f(r_s)\Big(1+\frac{L^2}{r_s^2}\Big)\Big(1-\frac{f(r)\Big(1+\frac{L^2}{r^2}\Big)}{f(r_s)\Big(1+\frac{L^2}{r_s^2}\Big)}\Big)^{1/2}}{f(r)\Big(1+\frac{L^2}{r^2}\Big)^{1/2}} + f(r_s)\frac{\Big(1+\frac{L^2}{r_s^2}\Big)}{\Big(1+\frac{L^2}{r^2}\Big)^{1/2}}
\nonumber
\\
&& + \frac{f(r_s)\Big(1+\frac{L^2}{r_s^2}\Big)\Big(1-\frac{f(r)\Big(1+\frac{L^2}{r^2}\Big)}{f(r_s)\Big(1+\frac{L^2}{r_s^2}\Big)}\Big)}{f^{2}(r)\Big(1+\frac{L^2}{r^2}\Big)^{1/2}}\Big]
\end{eqnarray}
and  for $r=r_s$ it is reduced to
\begin{equation}
F^{(\infty)}_U(r_s|r_s) =  \frac{\mathcal{L}}{4\pi r_s^2} f(r_s)\Big(1+\frac{L^2}{r_s^2}\Big)^{1/2}~,
\end{equation}
where it also yields finite quantity when one integrates it over the three space volume.

Finally, the flux in the Israel-Hartle-Hawking state near the horizon vanishes as
\begin{equation}
F^{(0)}_H(r|r_s) = 0.
\end{equation}
In this calculation, the matrix we have used is Eq. (\ref{Thh})
which is constant near the horizon. 
 That means the leading order of  Eq. (\ref{Thh}) gives us vanishing flux.
On the other hand, in the asymptotic infinite region, it is given by
\begin{eqnarray}
F^{(\infty)}_H(r|r_s) &=& \frac{1}{30~2^{12}\pi^2M^4}\frac{4E\Big(E^2-V_{eff}^2(r)\Big)^{1/2}}{3V_{eff}(r)\sqrt{f(r)}}
\nonumber
\\
&=&  \frac{1}{30~2^{12}\pi^2M^4}\frac{4f(r_s)\Big(1+\frac{L^2}{r_s^2}\Big)\Big[1-\frac{f(r)\Big(1+\frac{L^2}{r^2}\Big)}{f(r_s)\Big(1+\frac{L^2}{r_s^2}\Big)}\Big]^{1/2}}{3f^{3/2}(r)\sqrt{\Big(1+\frac{L^2}{r^2}\Big)}} 
\end{eqnarray}
which vanishes as $ F^{(\infty)}_H(r_s|r_s)=0$  at $r=r_s$,
 so that it is zero at the spatial infinity.

\section{Conclusion and discussion}
In the four-dimensional Schwarzschild black hole background, 
we have studied the energy densities and fluxes near the horizon
and asymptotic spatial infinity by using the 
trace anomaly of a conformally invariant scalar field in the three vacua.
In the Boulware state, both the energy density and flux are negative divergent 
at the horizon, and asymptotically vanish. In the Unruh state, 
the energy density is negative divergent at the horizon whereas it is positive finite
asymptotically. The flux in the Unruh state is always positive. 
For the Boulware and the Unruh states, the angular momentum due to 
the free-fall motion does not affect the energy density at the horizon.
However, in the Israel-Hartle-Hawking state, 
for $2M > L$, the energy density is negative at the horizon
whereas for $ 2M < L $ it can be positive, and 
the fluxes are zero at the horizon and the spatial infinity.
 
In particular, as for the flux in the Unruh state, which is alway positive. As was discussed,
the negative energy density near the horizon should be influx to reduce the 
black hole mass so that at first sight the flux
sign may be assumed to be negative. 
However, note that the energy density has already negative sign 
so that  the flux sign should be totally positive, which is similar
to the one-dimensional linear momentum case 
that the negative mass and negative velocity gives the positive sign.
Therefore, the ingoing negative energy density near the horizon toward the 
black hole and the out-going positive energy density at the spatial infinity
have the same sign.   

Unfortunately, we have just analytically discussed the limiting cases of the energy density
and flux instead of full region including the intermediate region
 without resort to numerical simulations which might 
disclose the full behaviors of them. However,
we obtained the important properties of the energy density and the flux
at the horizon. For instance, in the Unruh state describing 
the collapsing black hole, the energy density is negative near the horizon
and the positive asymptotically. It means that there exists at least one
free-fall position for the energy density to vanish. 
For the evaporating black hole, we can naturally imagine
that the black hole is surrounded by the negative energy density and
the positive energy density is located far away from the  horizon to the infinite
extent. The negative energy density
should be compensated with the positive black hole mass.
The positive energy is already separated from the black hole outside 
the horizon. If this positive energy carries information of black hole, then 
the negative energy carries negative information, which will result in
the unitary evolution of black hole as was discussed in Ref. 
\cite{Freivogel:2014dca}. 

Now, one might ask what the novel aspect of our calculation is. 
So we would like to emphasize some differences between the well-established works and the present calculation, and also discuss physical significances of the result. 
Hawking phenomenon [1] predicts that an observer at future infinity perceives an outgoing thermal flux of massless particles with a temperature. This effect is certainly observer dependent and one can not immediately say anything 
as to what different observers would measure. In this context, the important observation is due to Unruh [6]. The main result is that a static detector at a constant radial path detects thermal emission with a temperature which satisfies the Tolman equilibrium condition [26] just like a usual thermal bath. Next question arises: what will be for the detectors which are moving along the infalling geodesics? Since the Killing vectors are not tangent to these paths, one can not use the Tolman relation. Instead, since the acceleration for these frames is zero, it happens that there is no outgoing flux near the horizon following the Unruh effect. However, the above heuristic argument was challenged by ``firewall'' [18], where the principle of information theory suggests the presence of it very near the horizon. Unfortunately, till now there is no conclusive discussion in literatures to find the fate of the freely falling observers crossing the horizon. 

In our calculation, we clarified whether the freely falling observers perceive any radiation or not. 
In particular, it was found that the negatively "divergent" energy flows to the horizon rather than the high frequency
 outgoing modes at the horizon called the firewall [18], which has not yet been discussed in other literatures. 
In connection with black hole complementarity, the free-fall observer when crossing the horizon sees 
nothing special until he/she arrives at the coordinate invariant singularity. 
Therefore, the present calculation can modify the conventional black hole complementarity [3,4,5] 
since the huge negative energy density can be found when the observer is dropped very near the
horizon. Next, what needs to be answered is that why such a huge amount of energy density appears, 
which seems to be contradictory to the previous result giving vanishing result near the horizon in Ref. [6]. 
Actually, this is not inconsistent with the Unruh effect since the energy density largely consists of 
three components such as ingoing, outgoing, cross components of the energy-momentum tensors. 
Near the horizon, the outgoing  and cross components of energy-momentum tensors vanish,
so that there is no outgoing radiation following the Unruh effect. 
However, the non-vanishing ingoing energy-momentum tensor contributes to the infalling energy density, 
which implies that the ingoing negative energy can be found in the freely falling frame. 
Moreover, its magnitude is divergent at the horizon because the free-fall four velocity $u^a$,
 which plays a role of the vierbein $e^{a}_{\tau}$ to transform the fiducial coordinates to the locally flat coordinates
 where $\tau$ is the proper time in the locally flat spacetime, is divergent at the horizon as seen from Eq. (2).
 Note that the reason for this divergence is related to the gravitational time dilation. 
For instance, when the frame is dropped at rest very near the horizon, then the zeroth component 
of the four velocity of $u^{0}=dt/d\tau$ becomes very large since the coordinate time $dt$ is very large
while the proper time $d\tau$ is finite in freely falling frame. In these regards, we 
could see the nontrivial behaviors of the free-fall energy density near the horizon 
which are of relevance to modification of black hole complementarity including the firewall problem   
even in spite of the simplest context. This issue deserves further attention in this direction.

\vskip 5mm
\noindent
{\bf{Acknowledgements}}\\
\noindent
WK was indebted to Yongwan Gim for helpful comments
and would like to thank In Yong Park for exciting discussions.
This work was supported by the National Research Foundation of Korea(NRF) grant funded by the Korea government(MSIP) (2014R1A2A1A11049571). 
BRM was supported by a Lady Davis Fellowship at Hebrew University, 
by the I-CORE Program of the Planning and Budgeting Committee and the Israel Science Foundation 
(grant No. 1937/12), as well as by the Israel Science Foundation personal grant No. 24/12.


\begin{thebibliography}{99}
\bibitem{Hawking:1974sw} 
  S.~W.~Hawking,
  ``Particle Creation by Black Holes,''
  Commun.\ Math.\ Phys.\  {\bf 43}, 199 (1975)
  [Erratum-ibid.\  {\bf 46}, 206 (1976)].

\bibitem{Hawking:1976ra}
  S.~W.~Hawking,
 ``Breakdown of predictability in gravitational collapse,''
  Phys.\ Rev.\  D {\bf 14}, 2460 (1976).

\bibitem{Susskind:1993if}
  L.~Susskind, L.~Thorlacius and J.~Uglum,
  ``The stretched horizon and black hole complementarity,''
  Phys.\ Rev.\  D {\bf 48}, 3743 (1993)
  [arXiv:hep-th/9306069].


\bibitem{Stephens:1993an} 
  C.~R.~Stephens, G.~'t Hooft and B.~F.~Whiting,
  ``Black hole evaporation without information loss,''  
Class.\ Quant.\ Grav.\  {\bf 11}, 621 (1994)  [gr-qc/9310006].  


\bibitem{Susskind:1993mu}
  L.~Susskind and L.~Thorlacius,
  ``Gedanken experiments involving black holes,''
  Phys.\ Rev.\  D {\bf 49}, 966 (1994)
  [arXiv:hep-th/9308100].
  
\bibitem{Unruh:1976db}
  W.~G.~Unruh,
  ``Notes on black hole evaporation,'' 
 Phys.\ Rev.\ D {\bf 14}, 870 (1976).  

\bibitem{Ford:1993bw} 
  L.~H.~Ford and T.~A.~Roman,
  ``Motion of inertial observers through negative energy,''
  Phys.\ Rev.\ D {\bf 48}, 776 (1993)
  [gr-qc/9303038].

\bibitem{Callan:1992rs}
  C.~G.~Callan, Jr., S.~B.~Giddings, J.~A.~Harvey and A.~Strominger,
  ``Evanescent black holes,''
  Phys.\ Rev.\ D {\bf 45}, 1005 (1992)
  [hep-th/9111056].




\bibitem{Kim:2013caa} 
  W.~Kim and E.~J.~Son,
  ``Freely Falling Observer and Black Hole Radiation,'' 
 Mod.\ Phys.\ Lett.\ A {\bf 29}, 1450052 (2014)  [arXiv:1310.1458 [hep-th]].  

\bibitem{Boulware:1404}
  D.~G.~Boulware,
 ``Quantum Field Theory in Schwarzschild and Rindler Spaces,''  
 Phys.\ Rev.\ D {\bf 11} 1404 (1975).  

\bibitem{Israel:1976ur}
  W.~Israel,
 ``Thermo-field dynamics of black holes,''
  Phys.\ Lett.\ A {\bf 57}, 107 (1976).  


\bibitem{Hartle:1976tp}
  J.~B.~Hartle and S.~W.~Hawking,
 ``Path Integral Derivation of Black Hole Radiance,''
  Phys.\ Rev.\ D {\bf 13}, 2188 (1976).  


\bibitem{Christensen:1977jc} 
  S.~M.~Christensen and S.~A.~Fulling,
  ``Trace Anomalies and the Hawking Effect,'' 
   Phys.\ Rev.\ D {\bf 15}, 2088 (1977).  




\bibitem{Eune:2014eka} 
  M.~Eune, Y.~Gim and W.~Kim,
  ``Something special at the event horizon,''
  Mod.\ Phys.\ Lett.\ A {\bf 29}, 1450215 (2014)
  [arXiv:1401.3501 [hep-th]].

\bibitem{Park:2014mba} 
  I.~Y.~Park,
 ``Indication for unsmooth horizon induced by quantum gravity interaction,'' 
arXiv:1401.1492 [hep-th].  

\bibitem{Suprit} 
  M.~Smerlak and S.~Singh,
  ``New perspectives on Hawking radiation,''
  Phys.\ Rev.\ D {\bf 88}, 104023 (2013)
  [arXiv:1304.2858 [gr-qc]].

  
\bibitem{Freivogel:2014dca} 
  B.~Freivogel,
  ``Energy and Information Near Black Hole Horizons,''  
arXiv:1401.5340 [hep-th].  









\bibitem{Almheiri:2012rt} 
  A.~Almheiri, D.~Marolf, J.~Polchinski and J.~Sully,
  ``Black Holes: Complementarity or Firewalls?,''
  JHEP {\bf 1302}, 062 (2013)
  [arXiv:1207.3123 [hep-th]].


\bibitem{Braunstein:2013bra} 
  S.~L.~Braunstein, S.~Pirandola and K.~Zyczkowski,
  ``Better Late than Never: Information Retrieval from Black Holes,''
  Phys.\ Rev.\ Lett.\  {\bf 110}, no. 10, 101301 (2013).


    
  
        
\bibitem{Hutchinson:2013kka} 
  J.~Hutchinson and D.~Stojkovic,
  ``Icezones instead of firewalls: extended entanglement beyond the event horizon and unitary evaporation of a black hole,''
  arXiv:1307.5861 [hep-th].

\bibitem{Page:2013mqa} 
  D.~N.~Page,
  ``Excluding Black Hole Firewalls with Extreme Cosmic Censorship,''
  JCAP {\bf 1406}, 051 (2014)
  [arXiv:1306.0562 [hep-th]].
                          
\bibitem{Candelas}
    P. Candelas,
    ``Vacuum polarization in Schwarzschild spacetime,''
    Phys.\ Rev.\ D {\bf 21}, 2185 (1980).

\bibitem{Balbinot:1999vg} 
  R.~Balbinot, A.~Fabbri and I.~L.~Shapiro,
  ``Vacuum polarization in Schwarzschild space-time by anomaly induced effective actions,''
  Nucl.\ Phys.\ B {\bf 559}, 301 (1999)
  [hep-th/9904162].



\bibitem{Padmanabhan:2010zzb} 
  T.~Padmanabhan,
  ``Gravitation: Foundations and frontiers,''
  Cambridge, UK: Cambridge Univ. Pr. (2010) 700 p
 



\bibitem{Elster}
  T.~Elster,
  ``Vacuum polarization near a black hole creating particles,''
  Phys.\ Lett. A{\bf 94}, 205 (1983).


\bibitem{Tolman:1930zza} 
  R.~C.~Tolman,
  ``On the Weight of Heat and Thermal Equilibrium in General Relativity,''
  Phys.\ Rev.\  {\bf 35}, 904 (1930).








\end{thebibliography}
\end{document}